\documentclass[twocolumn]{aastex631}
\usepackage{graphicx} 
\usepackage{amsmath,bm}
\usepackage{mathtools}   
\usepackage{graphicx}
\usepackage[dvipsnames]{xcolor}
\usepackage{verbatim}
\usepackage{comment}
\usepackage{changepage}
\usepackage{float}
\usepackage{hyperref}
\usepackage{tabularx}
\usepackage{booktabs}
\usepackage[table]{xcolor}

\newcommand{\bfm}[1]{\mathbf{#1}}

\newcommand{\refsim}{{``\texttt{pulse\_n20\_ref}'' }}

\begin{document}
\title{
Damping
of Fast Radio Bursts 
in
the Inner Magnetospheres of Magnetars}
\author[0000-0001-6541-734X]{Siddhant Solanki}
\affiliation{Department of Astronomy, University of Maryland, 7901 Regents Drive, College Park, MD 20742, USA} 
\author[0000-0002-5349-7116]{Jens F. Mahlmann}
\affiliation{Department of Physics \& Astronomy, Wilder Laboratory, Dartmouth College, Hanover, NH 03755, USA}
\author[0000-0001-7801-0362]{Alexander Philippov}
\affiliation{Department of Physics, University of Maryland, 7901 Regents Drive, College Park, MD 20742, USA}
\affiliation{Physics Department, Stanford University, 382 Via Pueblo Mall, Stanford, CA 94305, USA}
\affiliation{
Kavli Institute for Particle Astrophysics and Cosmology, Physics and Astrophysics Building, 452 Lomita Mall, Stanford University, Stanford, CA 94305-
4085, USA}
\author[0000-0001-5660-3175]{Andrei M. Beloborodov}
\affiliation{Physics Department and Columbia Astrophysics Laboratory, Columbia University, 538  West 120th Street New York, NY 10027,USA}

\begin{abstract}
We investigate the propagation of fast radio bursts (FRBs) through magnetar magnetospheres. Previous work showed that, in the inner magnetosphere, GHz radio waves propagate as fast magnetosonic waves and undergo resonant three-wave interactions that transfer their energy into trapped Alfv\'en waves. Using three-dimensional force-free electrodynamics simulations, we demonstrate that FRBs would
excite Alfvénic fluctuations, leading to strong nonlinear
attenuation of the radio signal. In quiescent dipolar magnetospheres, the nonlinear decay stays efficient within $\sim10$--$100$ magnetar radii; charge starvation of the excited Alfv\'en waves stops the decay at larger radii. For FRBs propagating within relativistic magnetic outflows launched during magnetospheric eruptions, three-wave interactions remain efficient and constrain the escape radius to $\gtrsim10^2$--$10^3$ magnetar radii for luminous bursts. Our results confirm that nonlinear plasma processes strongly limit the escape of FRBs from the inner magnetospheres of magnetars.

\keywords{Waves -- Turbulence -- Magnetars(992) -- Instabilities -- Plasmas}
\end{abstract}

\section{Introduction} \label{section:Introduction}

FRBs are bright, millisecond-duration radio transients with observed frequencies spanning roughly $0.1$--$10~\mathrm{GHz}$ and isotropic-equivalent luminosities in the range $10^{38}$--$10^{44}~\mathrm{erg\,s^{-1}}$. Several thousand bursts have now been detected, originating from both repeating and apparently non-repeating sources \citep{chime_catalog_2021ApJS..257...59C,FRB_review_petroff_2019A&ARv..27....4P,FRB_review_petroff_2022A&ARv..30....2P,zhang_review_2023RvMP...95c5005Z}. Despite these abundant observations, the emission mechanism of FRBs is not established. A key observational clue comes from the Galactic magnetar SGR~1935+2154 \citep{2020Natur.587...54C}, whose contemporaneous radio and X-ray bursts demonstrate that at least some FRBs can be produced by magnetars---highly magnetized neutron stars with surface fields of $\sim 10^{14}$--$10^{15}~\mathrm{G}$ and spin periods of a few seconds \citep[see][for a review]{magnetar_review_2017ARA&A..55..261K}. 
Several theories have been proposed for how FRBs may
be generated in magnetar environments \citep[e.g.,][]{Lyubarsky2014:1401.6674v1,Beloborodov2017,Beloborodov2020,metzger_2019MNRAS.485.4091M,Kumar2020:2004.00644v1,Lu_kumar_frb_2020MNRAS.498.1397L,Lyubarsky_forced_reconnection2020:2001.02007v2,Sironi2021:2107.01211v1,Mahlmann_forced_reconnection2022:2203.04320v2,thompson_2023MNRAS.519..497T,Vanthieghem_levinson_ghz2024:2407.15076v1}. 

In one broad class of scenarios---the so-called \emph{magnetospheric} (``close-in'') models---the GHz radio emission originates within the inner magnetosphere, at distances of $\sim 10$--$10^2\,R_{\rm NS}$ from the star, where $R_{\rm NS}= 10^6~\mathrm{cm}$ is the neutron-star radius \citep{Kumar2020:2004.00644v1,Lu_kumar_frb_2020MNRAS.498.1397L,Zhang_IC2021:2111.06571v1,zhang_review_2023RvMP...95c5005Z,Qu2024:2404.11948v2}. In all such models, the extremely bright radio pulses must escape the strongly magnetized plasma of the inner magnetosphere, regardless of the specific mechanism responsible for generating them.

The inner regions of magnetar magnetospheres can be modeled using force-free electrodynamics (FFE), which describes the ultra--magnetized limit of magnetohydrodynamics in which plasma inertia and pressure are negligible. This approximation is justified by the extremely large plasma magnetization in the inner magnetosphere, $\sigma \equiv B^2/(4\pi \rho c^2) \gg 1$, where $B$ is the magnetic field strength, $\rho$ is the plasma mass density, and $c$ is the speed of light \citep{magnetar_review_2017ARA&A..55..261K}. Close to the magnetar, GHz waves propagate well below the local plasma frequency (see \citealt{golbraikh_2023ApJ...957..102G} and below) and therefore correspond to fast magnetosonic (FMS) waves of FFE. 

Two processes have been identified as constraining FRB propagation at these distances: scattering by individual particles, which occurs when the wave amplitude 
$E_0$
exceeds the background field 
$B_{\rm bg}$
\citep{Beloborodov_2022_PRL}, and resonant conversion into non-escaping Alfv\'en waves through three-wave processes,
which operate where $E_0<B_{\rm bg}$
\citep{golbraikh_2023ApJ...957..102G}. 
The latter work studied FRB conversion into Alfv\'en waves using kinetic equations of the weak-turbulence formalism with $E_0\ll B_{\rm bg}$ and assuming an isotropic wave distribution for simplicity. In this Letter, we extend the analysis in four ways: (i) we investigate the conversion process for a directed wave packet propagating through the background magnetosphere, (ii) we calculate the process using direct FFE simulations, (iii) we identify the range of radii where the MHD description holds and the conversion to Alfv\'en waves is enabled, and (iv) in addition to propagation through a static magnetosphere, we apply our simulation results to FRBs propagating inside relativistic ejecta of an erupting/exploding magnetosphere. We find a lower limit on the FRB production radius that applies even to explosion scenarios, which have $E_0<B_{\rm bg}$ at all radii.

\section{Plasma in the Magnetar Magnetosphere} \label{subsec:magnetar_magnetosphere}

In this Section, we summarize the justification for
treating FRBs as FMS waves in the FFE regime.
The magnetospheres of active magnetars are
filled with $e^{\pm}$ pairs produced in discharges 
sustaining magnetospheric electric currents \citep{electron_positron_flows_2013ApJ...777..114B}. The plasma density in the magnetosphere outside the pair-formation zone can be estimated as $n_{\rm bg} = \mathcal{N}/r^3$, where the dimensionless parameter $\mathcal{N}\sim 10^{37}$ is approximately constant with radius \citep{Beloborodov2020}. The plasma magnetization, $\sigma = B_{\rm bg}^2/(4\pi n_{\rm bg} m_e c^2)$, is 
huge
in the inner magnetosphere, where $B_{\rm bg} \sim \mu/r^3$ is the strength of the background magnetic field, and $\mu \sim 10^{33}\,{\rm G\,cm^3}$ is the magnetic moment of the magnetar.
We estimate the plasma frequency as
\begin{equation}
    \nu_p = \sqrt{\frac{n_{\rm bg} e^2}{\pi m_e}} \sim 10^{12} \mathcal{N}_{37}^{1/2}r_7^{-3/2} \text{ Hz}.
\end{equation}
where we employ notation $q=10^xq_{,x}$ in CGS units. Thus, for plasma densities expected in quiescent magnetospheres, $\nu_p$ is well above the FRB frequency, $\nu_{\rm FRB}\sim 1\,{\rm GHz}$, within $\sim 10^3\,R_{\rm NS}$ \citep{golbraikh_2023ApJ...957..102G}. The strength of an
electric field perturbation, ${E}_{\rm F0}$, in an FRB signal can be estimated from a fiducial isotropic FRB luminosity, $L\sim cr^2E^2_{\rm F0}/2$. Relative to the background field $B_{\rm bg}$,
it increases progressively as the FRB propagates away from the star:
\begin{equation}
            \frac{{E}_{\rm F0}}{{B}_{\rm bg}} \sim 3 \times 10^{-3}   
            L^{1/2}_{43}r^2_7 \label{eq:relative_strength},
\end{equation}
and $E_{\rm F0}$ approaches $B_{\rm bg}$ at distances $\sim {\rm few} \times 10^2 R_*$ \citep{beloborodov_khz_2021ApJ...922L...7B}. 

Thus, in the inner magnetar magnetosphere, typical FRB frequencies lie well below the plasma frequency, and the amplitude of their electromagnetic perturbations is much smaller than the background magnetic field. The plasma is also highly magnetized, $\sigma \gg 1$, 
and the Larmor frequency of particles far exceeds the GHz wave frequency.
Consequently, GHz-frequency FRBs can be described as linear FMS waves
in the MHD fluid. The fluid also sustains the excited Alfv\'en waves, as long as charge starvation is avoided; this condition will be discussed in section~\ref{starve} below.

\section{Three-wave interactions} \label{section:theory}

Consider an FMS pulse propagating in the inner magnetosphere with wave vector $\boldsymbol{k}_1$, making an angle $\theta$ with the background magnetic field $\boldsymbol{{B}}_0$ in the frame of the magnetar. The pulse electromagnetic fields follow the FMS eigenmode polarization and may be written as $\boldsymbol{E}_{\rm F} = f(\xi)\left(\boldsymbol{\hat{k}}_{1,\perp} \times \boldsymbol{\hat{B}}_0\right), \boldsymbol{B}_{\rm F} = f(\xi)\left(\cos\theta\, \boldsymbol{\hat{k}}_{1,\perp} - \sin\theta\, \boldsymbol{\hat{B}}_0\right)$, where $\xi \equiv \boldsymbol{\hat{k}}_1\cdot\boldsymbol{r}$ is the coordinate along the direction of propagation, $f(\xi)$ describes the spatial profile of the pulse, and $\boldsymbol{\hat{k}}_{1,\perp}$ is the normalized component of $\boldsymbol{k}_1$ perpendicular to $\boldsymbol{\hat{B}}_0$. We perform a Lorentz transformation to a frame moving with velocity $\boldsymbol{\beta}_{\rm boost} = \cos\theta\, \boldsymbol{\hat{B}}_0$, such that in the boosted frame the pulse propagates perpendicular to the background field, with group velocity along $\boldsymbol{\hat{k}}_1' \equiv \boldsymbol{\hat{k}}_{1,\perp}$. In this frame, the pulse fields transform to $\boldsymbol{E}_{\rm F}' = f(\xi')\, \sin\theta\, \left(\boldsymbol{\hat{k}}_1' \times \boldsymbol{\hat{B}}_0\right), 
\boldsymbol{B}_{\rm F}' = - f(\xi')\, \sin\theta\, \boldsymbol{\hat{B}}_0,$
where $\xi' \equiv \boldsymbol{\hat{k}}_1'\cdot\boldsymbol{r}$. We carry out the subsequent analysis in this boosted frame and hereafter omit primes on all quantities.

In the absence of other perturbations and neglecting curvature of the background field (appropriate for radio wavelengths), the pulse electromagnetic fields are exact solutions of the FFE equations and propagate as $f(\xi-ct)$ in time, so one could naively expect them to escape freely. However, in realistic magnetospheres the FMS pulse (F) can encounter low-amplitude Alfvénic waves (A) with amplitudes $|\boldsymbol{E}_{\rm A}|\ll|\boldsymbol{E}_{\rm F}|$. These perturbations may be parametrically amplified through three-wave resonant interactions such as $F\to F+A$ and $F\to A+A$ \citep{golbraikh_2023ApJ...957..102G,lyubarsky_review_frb_2021Univ....7...56L,Li_beloborodov_2019ApJ...881...13L,alex_chen_2024arXiv240406431C,companion}. A resonant triad satisfies the four-vector resonance condition
$
\delta k^\mu \equiv k_1^\mu - k_2^\mu - k_3^\mu = 0,
$
where $k_i^\mu=(\omega(\boldsymbol{k}_i)/c,\;\boldsymbol{k}_i)$ for $i=1,2,3$ (the incoming FMS and the two resonantly driven modes). The FMS and Alfvén-wave dispersion relations are $
\omega_{\rm F} = c|\boldsymbol{k}|, \,
\omega_{\rm A} = c\,|\boldsymbol{k}\cdot\hat{\boldsymbol{B}}_0|,
$
which constrain the allowed resonant wavevectors entering the triad.

We first consider the propagation of a monochromatic FMS-wave pulse of wavelength $\lambda_1 \equiv 2\pi k_1^{-1}$ and envelope size $\Delta$. We adopt a Cartesian coordinate system in which the pulse propagates along the $x$-axis, so that $\boldsymbol{k}_1 = (k_1,0,0)$ with $k_1>0$, and the background field is $\boldsymbol{B}_0 = (0,0,B_0)$. When an FMS wave interacts with an AW with wavevector $\boldsymbol{k}_2$, the AW charge density $\rho_{\rm A}$ experiences an $\boldsymbol{E}_{\rm F}\times\boldsymbol{B}_0$ drift in response to the FMS electric field, producing a nonlinear current $\boldsymbol{J}_{\rm NL} = (\rho_{\rm A} c)\, \boldsymbol{E}_{\rm F}\times\boldsymbol{B}_0 / |\boldsymbol{B}_0|^2$. In addition, the superposition of the FMS and AW fields induces second-order fields $\boldsymbol{E}^{(2)} = -\hat{\boldsymbol{z}}(\boldsymbol{E}_{\rm F}\cdot\boldsymbol{B}_{\rm A})/|\boldsymbol{B}_0|$ and $\boldsymbol{B}^{(2)} = -c\int_0^t (\nabla\times\boldsymbol{E}^{(2)})\,{\rm d}t'$, which arise to maintain the FFE condition $\boldsymbol{E}\cdot\boldsymbol{B}=0$. Both $\boldsymbol{J}_{\rm NL}$ and $\nabla\times\boldsymbol{B}^{(2)}$ contain Fourier components with angular frequencies $\omega = \omega_1 \pm \omega_2$ and wavevectors $\boldsymbol{k} = \pm(\boldsymbol{k}_1 \pm \boldsymbol{k}_2)$. These terms can resonantly drive the exponential growth of a second AW with wavevector $\boldsymbol{k}_3$, provided the three-wave matching condition $k_3^{\mu} = k_1^{\mu} - k_2^{\mu}$ is satisfied. For an FMS wave propagating along $x$, this requires $k_{2z} = -k_{3z} = \pm k_1/2$, while $k_{2x}$ and $k_{2y}$ remain unconstrained. Consequently, AWs with large perpendicular wavenumbers $k_{2\perp}$ can participate in the interaction, allowing for a large phase space for the decay of FMS waves
 \citep{lyubarsky2021emission}. Additionally, another FMS wave can be sourced via the $F\to F+A$ process.

In the limit \(k_{2\perp}, k_{3\perp} \gg k_1\), the growth rate, $\gamma_{\rm NL}$, for the excitation of AWs becomes only weakly sensitive to the specific values of \(\boldsymbol{k}_2\) and \(\boldsymbol{k}_3\) \citep{golbraikh_2023ApJ...957..102G, companion}:
\begin{equation}
\gamma_{\rm NL} \approx \frac{\omega}{2} \frac{|\boldsymbol{E}_{\rm F0}|}{|\boldsymbol{B}_0|} ,
\end{equation}
where \(\boldsymbol{E}_{\rm F0}\) is the initial electric field amplitude of the FMS wave. Since the group velocity of Alfvén waves is aligned with the background magnetic field and perpendicular to the direction of pulse propagation, the interaction time for resonant processes is limited to \(\sim \Delta/c\). The pulse propagation is thus controlled by the number of e-foldings of the instability that occur during the light-crossing time of the pulse:

\begin{equation}
\eta = \gamma_{\rm NL} \frac{\Delta}{c}  \approx \pi N_{\lambda}  \frac{|\boldsymbol{E}_{\rm F0}|}{|\boldsymbol{B}_0|},
\label{eq:e-fold}
\end{equation}
where \(N_\lambda\) is the number of wavelengths contained in the pulse.  When \(\eta \gg 1\), the AWs grow to amplitudes comparable to the FMS wave within the pulse, leading to efficient conversion of the initial FMS energy into AWs that cannot escape the magnetosphere. 
This conversion occurs in wave packets with $N_\lambda \gg 1$, as observed in FRBs.
Conversely, for \(\eta \lesssim 1\), the interaction remains weak, as the AWs do not have sufficient time to reach amplitudes comparable to that of FMS waves. Short pulses with \(N_\lambda \sim \text{a few}\) typically fall into this regime unless \(|\boldsymbol{E}_{\rm F0}| \sim |\boldsymbol{B}_{0}|\).  

For broadband pulses, \(\Delta \nu \sim \nu\), the interaction rate is reduced \citep{golbraikh_2023ApJ...957..102G}, scaling as $\gamma_{\rm NL} \sim \omega \left({|\boldsymbol{E}_{\rm F0}|}/{|\boldsymbol{B}_0|} \right)^2$. Consequently, broadband pulses must be longer to allow efficient conversion into AWs. We test the propagation of both monochromatic and broadband pulses using three-dimensional FFE simulations, which are described in the following section.

\section{FFE simulations}
\label{section:theory}

\subsection{Setup}
We perform three simulations (``\texttt{pulse\_n20\_ref}'',  ``\texttt{pulse\_mixed}'' and ``\texttt{pulse\_n6}'') using the FFE module \citep{CarpetCode:web,Jens_code_1_2021A&A...647A..57M,Jens_code_2_2021A&A...647A..58M} in the \textsc{Einstein Toolkit} (ET) framework \citep{Loeffler2012}. The simulations employ Cartesian coordinates $(x,y,z)$ with periodic boundary conditions along $\pm y$ and $\pm z$, absorbing boundary conditions at the $\pm x$ edges, and a uniform background magnetic field $B_0=1$ in the $+z$ direction. The domain extends in $(0,5L)$ range along $x$ and $(0,L)$ in both $y$ and $z$. We initialize an FMS pulse propagating in the $+x$ direction with $E_{y,{\rm F}} = B_{z,{\rm F}} = f(x)$. For narrowband simulations (``\texttt{pulse\_n20\_ref}'' and ``\texttt{pulse\_n6}'') we choose $f(x) = A \sin[k_1(x-x_0)] \exp[-((x-x_0)/\Delta)^6]$, with $A=0.3$, $x_0/L=1$, $k_1L/(2\pi)=14$, and we vary $\Delta$, which determines $N_\lambda$ and sets $\eta$. The broadband simulation (``\texttt{pulse\_mixed}'') is initialized with $N=20$ FMS modes, where $f(x) = \tanh(g(x))$, and $g(x) = (A/\sqrt{N}) \sum_{n=1}^{N} \sin[k_n(x-x_0) + \phi_{k_n}] \exp[-((x-x_0)/\Delta)^6]$, where $(k_n-k_0)L/2\pi=2n$, $k_0L/(2\pi)=20$, $\Delta/L = 0.75$, and $\phi_{k_n}$ is a random phase in $[0,2\pi)$. The $\tanh(g(x))$ transformation caps large values of $g(x)$. In all simulations, the initial FMS pulse is well resolved, with at least 18 cells per FMS wavelength. To model nonlinear wave interactions in a realistic background, the simulations include a low-amplitude spectrum of AWs with total energy density $U_{\rm A}/U_0 = 5\times10^{-5}$, where $U_0 \equiv B_0^2/(4\pi)$.  Table~\ref{tab:sims} summarizes the parameters of our simulations.

\begin{table}[b]
\caption{\label{tab:sims} Summary of simulations. $N_{\lambda}$ denotes the number of wavelengths in the FMS pulse with amplitudes greater than $A/2$. For the broadband ``\texttt{pulse\_mixed}'' simulation, all parameters are estimated using the pulse’s central frequency.}
\begin{ruledtabular}
\begin{tabular}{lccccc}
Simulation & Resolution & $\Delta$ & $N_{\lambda}$ &  $\eta$  \\
\hline
\texttt{pulse\_n20\_ref} & $2560\times 512^2$ & 0.75 & 20 & 19\\
\texttt{pulse\_mixed}    & $5120\times 512^2$ & 0.75 & 57 & 11 \\
\texttt{pulse\_n6}       & $1280\times 256^2$ & 0.25 & 6  & 5.6\\
\end{tabular}
\end{ruledtabular}
\end{table}

\begin{figure*}[]
    \centering
    \includegraphics[width=0.95\textwidth]{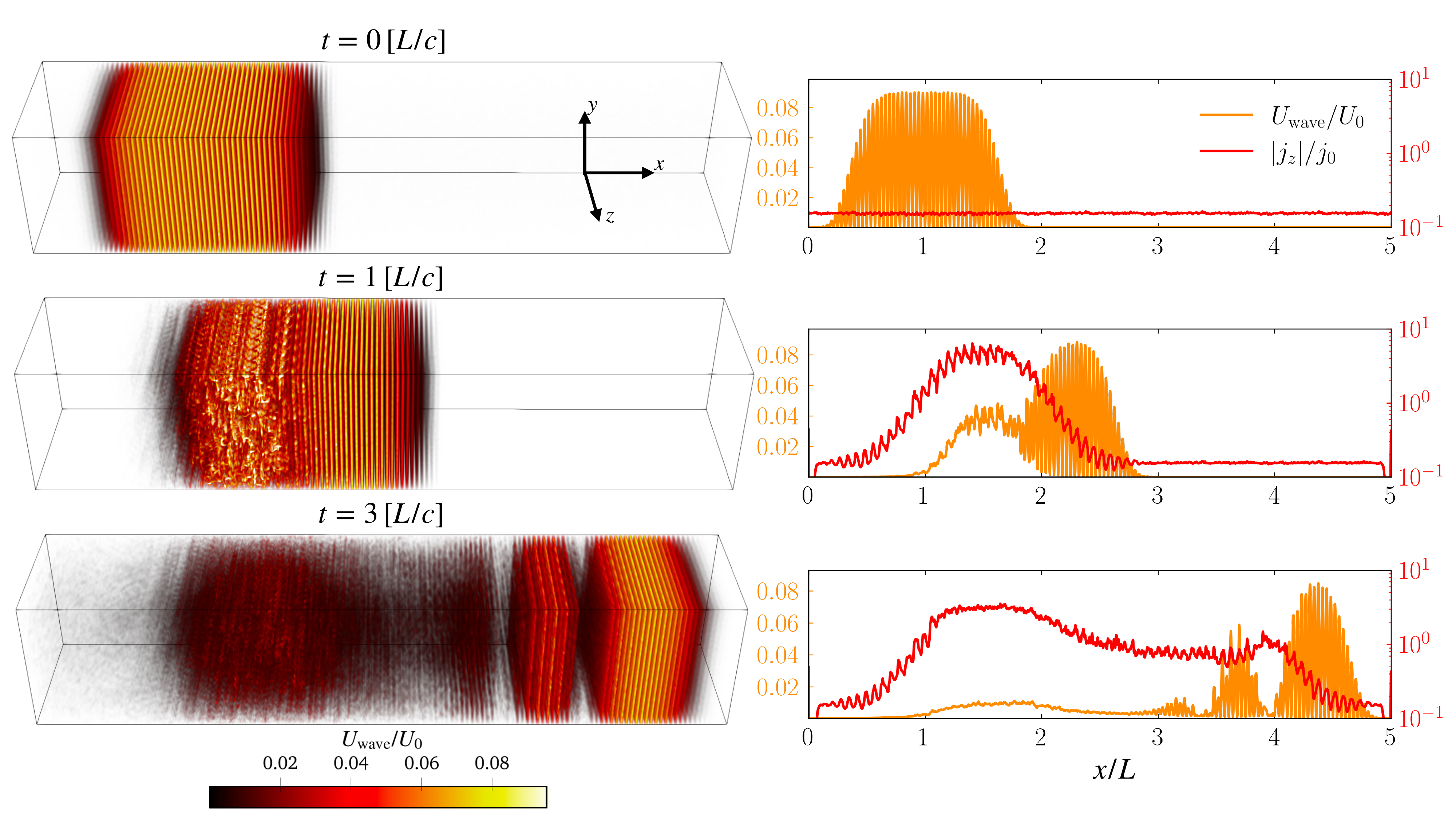}
    \caption{Evolution of the fast magnetosonic pulse in the \refsim\ simulation. The left column shows the volume renderings of the electromagnetic wave energy density, $U_{\rm wave}/U_0$, and the right column shows the $x$-profiles of the $y$--$z$ plane–averaged electromagnetic energy density (orange) and field-aligned component of the current, $|j_z|/j_0$ (red), as indicated by the corresponding colors on the $y-$axes.}
    \label{fig:energy_density_rendering}
\end{figure*}

\begin{figure*}[]
    \centering
    \includegraphics[width=0.95\textwidth]{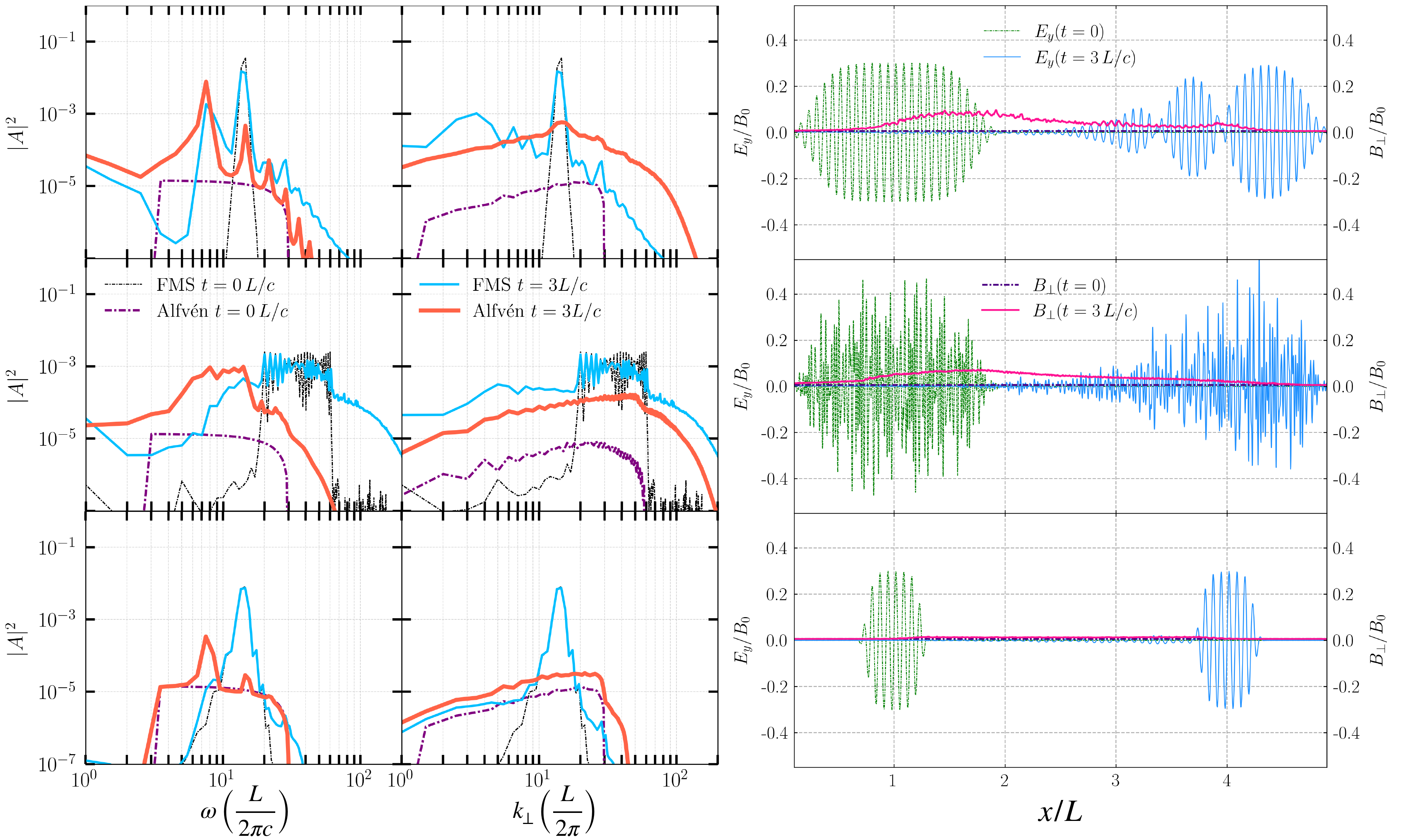}
    \caption{Evolution of the FMS pulse in simulations \refsim (top row), \texttt{pulse\_mixed} (middle row), and \texttt{pulse\_n6} (bottom row). The left column shows the frequency-space and $k_{\perp}$-space power spectra of the FMS and Alfvén waves at $t=0$ and $3\,L/c$. Resonant Alfvén waves grow at frequencies $\omega \approx k_1 c/2$ and their harmonics, corresponding to a broad $k_{\perp}$ spectrum. The different colors in the right column represent $y$--$z$ plane averages of $E_y/B_0$ and $B_{\perp}/B_0$ at times $t=0$ and $3\,L/c$. The electric field of the fast magnetosonic wave is polarized along the $y$ direction, while the excited Alfvén waves produce an increase in the magnetic-field component transverse to the background field, $B_{\perp}$.}
    \label{fig:spectrum}
\end{figure*}

\subsection{Results} Figure~\ref{fig:energy_density_rendering} shows the evolution of the FMS pulse in our reference simulation, \refsim. As the pulse propagates in the $+x$ direction, the first few wavelengths at its leading edge remain unperturbed, while the electromagnetic energy in the tail rapidly decreases. This behavior is illustrated by the electromagnetic energy density, $U_{\rm wave} \equiv (|\bfm{E}|^2 + |\bfm{B}-\bfm{B}_0|^2)/(8\pi)$, shown at various times in both columns.

The decay process is accompanied by the excitation of short-wavelength waves. These waves carry field-aligned conduction currents, $j_z$, along the background field direction, $\boldsymbol{\hat{z}}$, and are left behind as the leading edge of the FMS pulse propagates forward. This behavior is illustrated in the right column of Figure~\ref{fig:energy_density_rendering}, which shows the $y$--$z$ plane–averaged values of $U_{\rm wave}/U_0$ and the field-aligned current density, $|j_z|$, normalized by $j_0 \equiv c/(4\pi)\, A k_1$, at the corresponding times. Thus, the excited waves are consistent with Alfvén waves, whose group velocity is directed along the background magnetic field.

We next quantify the decay process by performing a Fourier analysis of the electromagnetic fields. Figure~\ref{fig:spectrum} shows the frequency spectra, $A_{\omega}$, the perpendicular wavenumber spectra, $A_{k_{\perp}}$, and the $y$--$z$ plane averages of $E_y/B_0$, which trace the polarization of the FMS wave, and $B_{\perp}/B_0$, indicative of AWs, for the \refsim, ``\texttt{pulse\_mixed},'' and ``\texttt{pulse\_n6}'' simulations at times $t=0$ (dashed) and $t=3L/c$ (solid). The spectral amplitudes are decomposed into FMS and AW eigenvectors. As seen in the solid orange lines in the left panels, resonant Alfvén waves with $\omega_{\rm A} \approx \omega_{\rm F}/2$ grow exponentially by several orders of magnitude in the \refsim and ``\texttt{pulse\_mixed}'' runs, whereas Alfvén wave growth is only marginal in ``\texttt{pulse\_n6}'', consistent with the small number of e-foldings expected for short pulses. The generated AWs exhibit a broad spectrum in $k_{\perp}$ space, extending to the grid scale where dissipation becomes efficient, $k_{\perp} L / 2\pi \approx 10^2$. In the ``\texttt{pulse\_mixed}'' simulation, only the higher-frequency components of the FMS pulse, for which $\eta \gtrsim {\rm few}$, efficiently transfer their energy to Alfvén waves. In addition, we observe spectral broadening of the FMS waves toward large $k_{\perp}$, driven by higher-order nonlinear processes. A notable ``echo'' appears in the $E_y$ profile in the \refsim simulation, where the leading edge of the original pulse is followed by a sequence of shorter sub-pulses of progressively smaller amplitude. These sub-pulses share the same polarization as the original FMS pulse and are therefore identified as FMS waves generated by resonant interactions of Alfvén waves, with their amplitudes decreasing as the Alfvén-wave spectrum broadens and the efficiency of the $A + A \to F$ process declines.

Our analysis confirms that FMS pulses of sufficiently large amplitude and duration, allowing for a large number of e-foldings of the resonant processes, efficiently convert their energy into Alfvén waves that remain trapped along the background magnetic-field lines.


\section{Application to Fast Radio Bursts}

\begin{figure}
    \centering
    \includegraphics[width=0.5\textwidth]{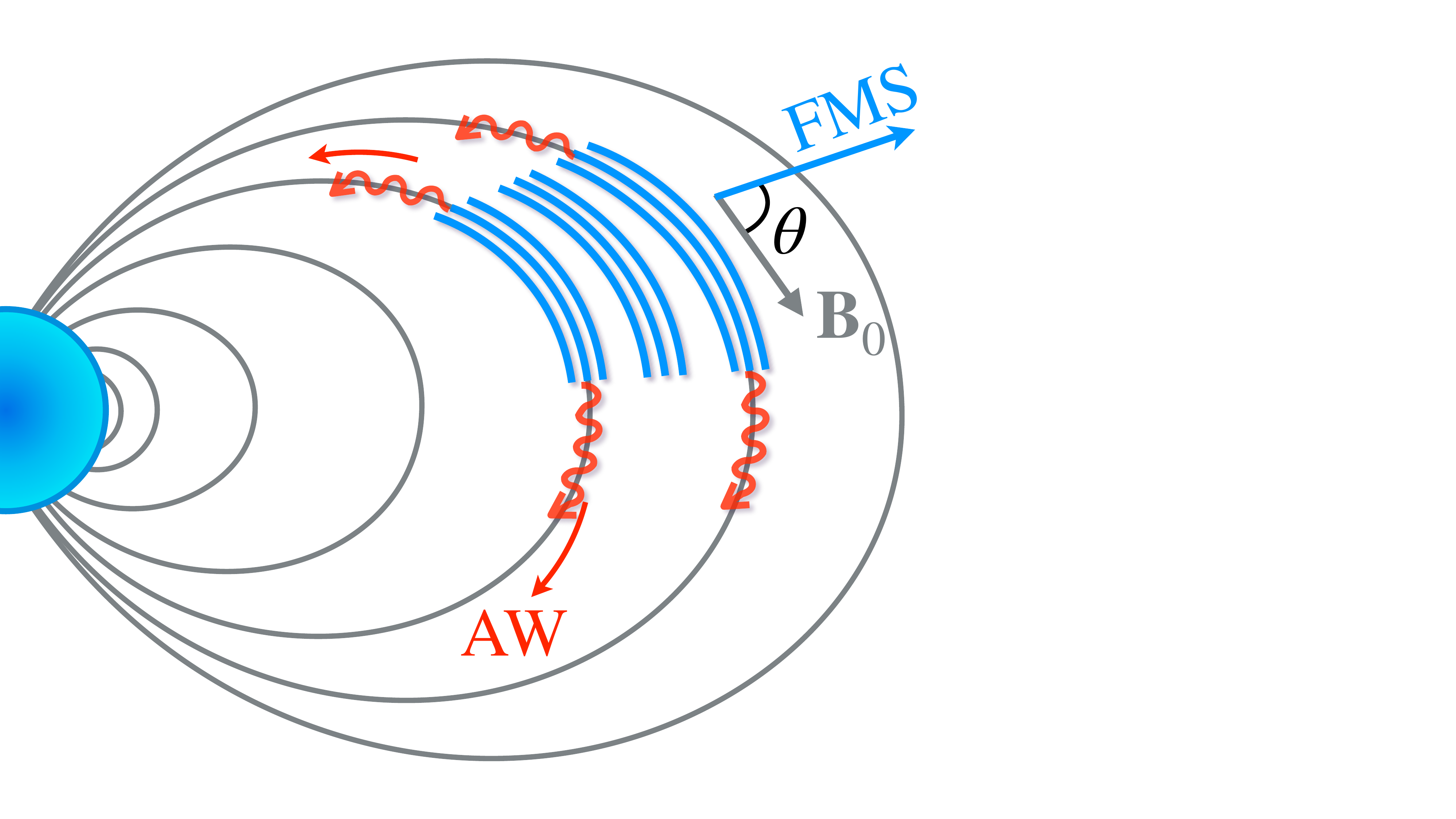}
    \includegraphics[width=0.5\textwidth]{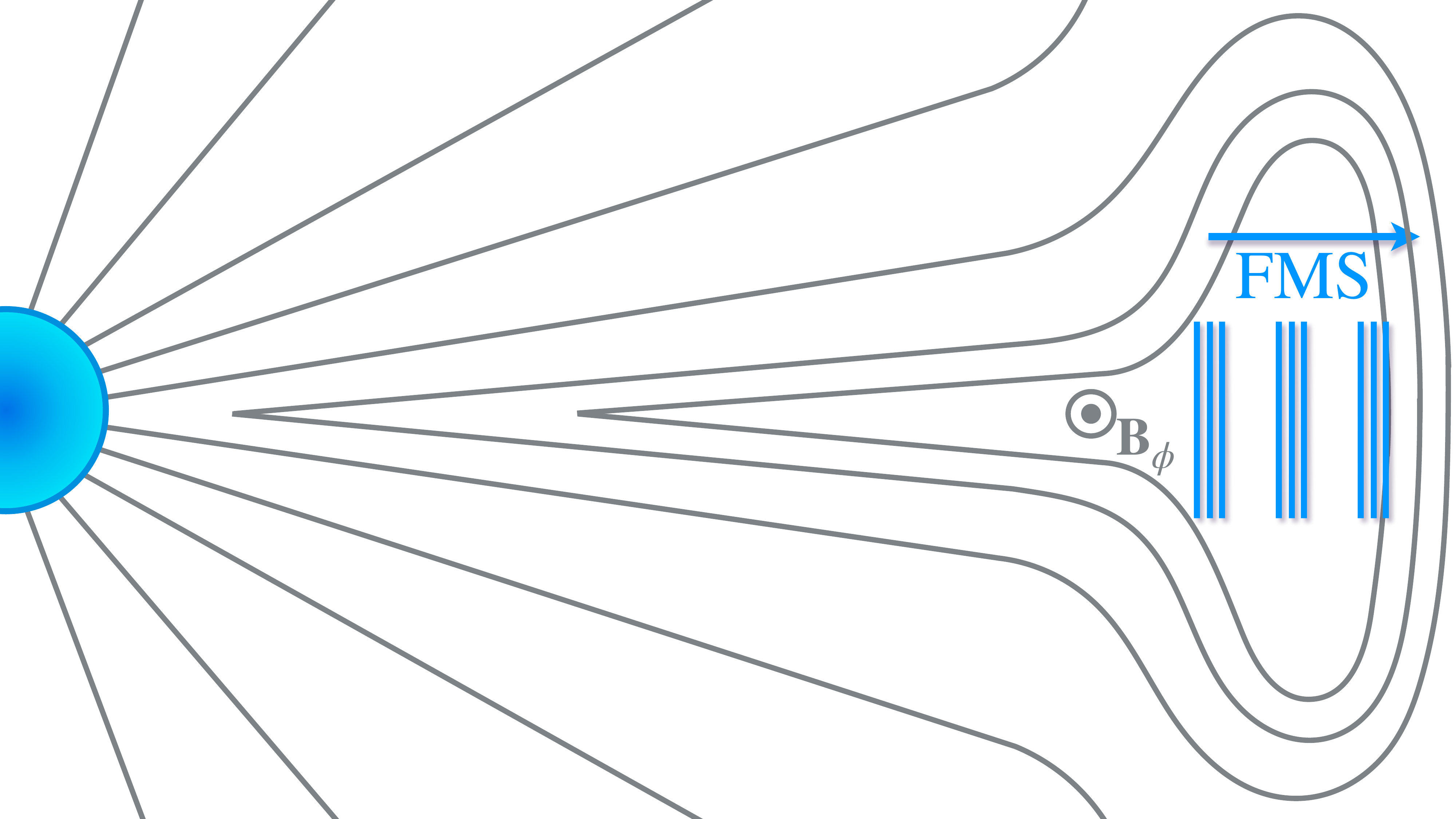}
    \caption{Illustration of two FRB propagation scenarios: in a quiescent magnetosphere (top panel) and riding on an outward-propagating magnetic pulse ejected by a magnetar magnetosphere (bottom panel).}
    \label{fig:FRB_propagation}
\end{figure}

We now
discuss the application of our 
results
to FRBs. We consider 
two scenarios: FRB
propagation in quiescent, unperturbed magnetospheres and propagation atop an escaping magnetic pulse ejected by an overtwisted magnetosphere, as illustrated in Figure~\ref{fig:FRB_propagation}.

\subsection{Propagation in quiescent magnetospheres}
The number of e-foldings of the parametric decay instability for radio signals with FRB-equivalent luminosities can be estimated as a function of distance from the magnetar. Below, we estimate the efficiency of the decay process for a broadband FRB, $\Delta \nu \gtrsim \nu_{\rm FRB}$, since this case imposes less stringent conditions for wave escape. For a broadband spectrum, the interaction rate for oblique waves is $\gamma_{NL}\sim \omega_{\rm FRB} ({B}_{\rm F,0}/B_{\rm bg})^2\sin^4\theta$. For propagation through an unperturbed dipolar magnetosphere, the relative amplitude of the FMS waves, $|{B}_{\rm F,0}|/B_{\rm bg}$, increases with distance, resulting in an enhanced probability of resonant interactions.


Let us now estimate the radius, $r_{\rm turb}$, in the magnetosphere at which nonlinear processes set in. For an FMS wave propagating at an angle $\theta$ with respect to the background magnetic field, the 
interaction timescale
is determined by the residence time of excited Alfv\'en waves inside the FMS wave packet:
\begin{equation}
    t_{\rm max} \sim \frac{\Delta t_{\rm FRB}}{1-\cos\theta},
\end{equation}
where $\Delta t_{\rm FRB}\sim 10^{-3}\,{\rm s}$ is the duration of the FRB pulse, and the factor $(1-\cos\theta)$ arises from the difference in group velocities between FMS waves and AWs. 

The growth of Alfv\'en waves depends exponentially on the parameter
\begin{equation}
\eta=\gamma_{NL}(r_{\rm turb})\, t_{\rm max} \sim \frac{2\pi N_{\lambda}({B}_{\rm F,0}/B_0)^2\sin^4\theta}{1-\cos\theta},
\end{equation}
where
$N_\lambda = \nu_{\rm FRB}\Delta t_{\rm FRB} \sim 10^{6}$ is the number of wavelengths in the FRB pulse. The process significantly depletes the FMS wave packet if $\eta$ reaches
$\sim \ln({E_{\rm F,0}}/E_{\rm AW, noise}) \sim 10$ 
---
the number of e-foldings to grow the AWs from the noise level, $E_{\rm AW, noise}$, to amplitudes comparable to that of 
the
FRB.
For a quiescent dipolar magnetosphere, this condition yields
\begin{align}
\frac{r_{\rm turb}}{10R_{\rm NS}} \, \approx\, & 0.7\,  \frac{(1-\cos\theta)^{1/4}}{\sin \theta}\eta^{1/4}_1\nonumber\\
&\times\Delta t_{\rm FRB,-3}^{-1/4} 
\nu_9^{-1/4} L_{43}^{-1/4}.
\label{eq:critical radius}
\end{align}

Near the magnetic axis, i.e., for small propagation angles $\theta$ with respect to the magnetic field, the scaling of the critical radius, $r_{\rm turb}/R_{\rm NS} \sim \theta^{-1/2}$, is similar to that found for the ``monster'' shock formation \citep{monster_shock_Beloborodov2022:2210.13509v3}. We also note that for a truly narrowband signal, $\gamma_{\rm NL}\sim \sin^2\theta$, which implies $r_{\rm turb}/R_{\rm NS} \sim \theta^{0}$. In this case, even waves propagating close to the magnetic axis can be efficiently reprocessed via nonlinear three-wave interactions.

\subsection{The role of charge starvation for AWs}
\label{starve}

Our treatment of FMS wave decay assumes that FFE captures all relevant dynamics and nonlinear interactions. As discussed above, modeling FRBs as FMS waves is well justified within $\sim 10^2\,R_{\rm NS}$ for plasma densities expected in quiescent magnetospheres. The high-$k_{\perp}$ AWs produced by FMS decay, however, require substantial current densities, $j_{\rm AW} \approx (c/4\pi)\, k_{\perp} E_{\rm AW}$, and may encounter charge starvation when $j_{\rm AW} / (n e c) \gtrsim 1$ \citep[e.g.,][]{Beloborodov2023:2307.12182v2,cs_2022PhRvL.128g5101N}. The amplitude of AWs at which charge starvation becomes important, $E_{\rm AW, st}$, normalized to the electric field of the FRB signal, is given by
\begin{align}
\nonumber
\psi_{\rm st} & =\frac{E_{\rm AW, st}}{E_{\rm F0}}
 \sim
\frac{4\pi n_{\rm bg} e r}{k_{\perp} (2L_{\rm FRB}/c)^{1/2}} \\
 &\approx
0.2 \left(\frac{r}{10R_{\rm NS}}\right)^{-2}
L^{-1/2}_{43}\nu^{-1}_{\rm{FRB},9} \, {\mathcal{N}_{37}},
\end{align}
for AWs with $k_{\perp} \sim k_{\parallel, \rm A} \sim k_{\parallel, \rm F}/2\sim \omega_{\rm{FRB}}/2c$. Therefore, beyond $\sim 10R_{\rm NS}$, AWs excited via three-wave decay become marginally charge-starved before their amplitudes can approach that of the FMS wave. At this stage, they develop an electric field parallel to the background magnetic field and can no longer be described within the FFE framework. Consequently, charge starvation 
tends to
inhibit the decay of FMS waves. 

As argued by \cite{golbraikh_2023ApJ...957..102G}, 
pair creation prevents charge starvation and promotes the FMS decay. Indeed,
the electric fields 
induced in
charge-starved AWs 
would
accelerate electrons and positrons, which then emit curvature photons. In the presence of a strong background magnetic field, these photons can produce electron--positron pairs via the $\gamma - B$ process. This pair-production mechanism can replenish the plasma density to levels sufficient to prevent charge starvation of the excited waves, thereby allowing the decay of the FMS wave to proceed.

Let us 
examine this process in more detail. The parallel electric field in excited charge-starved AWs accelerates electrons and positrons to Lorentz factors 
\begin{equation}
    \gamma\sim a_{\rm st}\sim 4\left(\frac{\omega_p}{\omega_{\rm FRB}}\right)^2 \approx 3\times 10^6r_7^{-3}\nu^{-2}_{\rm{FRB},9}  {\mathcal{N}_{37}},
\end{equation}
where $a_{\rm st}\sim eE_{\rm AW}/(m_e c \omega_{\rm AW})=(\omega_p/\omega_{\rm AW})^2$ is the $a$-parameter of charge-starved AWs \citep{Beloborodov2023:2307.12182v2}, and $\omega_{\rm AW}\sim \omega_{\rm FRB}/2$. The accelerated particles stream along the field lines, while also performing the $\bfm{E} \times \bfm{B}$ drift. The time-varying fields of the FRB induce an oscillatory $\bfm{E} \times \bfm{B}$ drift, which imparts a curvature in the particle trajectories with a radius $r_c \sim (c / \omega_{\rm FRB}) (B_{\rm bg}/E_{\rm F0})$. Thus, particles emit curvature radiation with a characteristic photon energy $\hbar \omega_c \approx (3/2)\hbar\, (c / r_c)a^3_{\rm st}$, which can produce secondary electron-positron pairs. However, as the 
FRB propagates outward,
the Lorentz factors of particles accelerated by the 
generated
marginally charge-starved AWs 
drop as $\gamma\propto r^{-3}$.
This leads to lower-energy photons and a reduced probability of pair production. Efficient $\gamma-B$ pair production is expected only for $\chi_{q} \gtrsim 0.1$, where $\chi_{q}$ characterizes the probability of photon conversion in a strong magnetic field:
\begin{align}
\begin{split}
\chi_{q} 
&= \frac{\hbar \omega_c}{m_e c^2} \frac{B_{\rm bg}}{B_c}
\approx 96 \frac{\hbar \omega_{\rm FRB}}{m_e c^2} \frac{E_{\rm FRB}}{B_c}
\Bigl(\frac{\omega_p}{\omega_{\rm FRB}}\Bigr)^6 \\
&\approx 0.1 \left(\frac{r}{35 R_{\rm NS}}\right)^{-10} L^{1/2}_{ 43}\nu^{-5}_{\rm{FRB},9}{\mathcal{N}^3_{37}},
\end{split}
\end{align}
where $B_c \approx 4.4 \times 10^{13}\,{\rm G}$ is the quantum critical field. 
Hence, pair creation via the $\gamma-B$ process is suppressed at $r\gtrsim 35 R_{\rm NS}$.

Additional pairs 
are, however,
produced via $\gamma$–$\gamma$ collisions. The radiation energy density is comparable to that of charge-starved AWs, $\sim E^2_{\rm AW,st}/4\pi$. The corresponding radiative compactness parameter can be estimated as
\begin{equation}
\ell=\frac{\sigma_T E^2_{\rm AW,st} r}{4\pi m_e c^2}
\sim 2\left(\frac{r}{100R_{\rm NS}}\right)^{-5}\nu^{-2}_{\rm FRB,9}\mathcal{N}^2_{37},
\end{equation}
which is large within $\sim 100R_{\rm NS}$. Therefore, efficient pair production is expected provided the curvature photons are sufficiently energetic, 
i.e. at least a fraction of photons have energies above $m_ec^2$. For the typical energy of the curvature photons $\hbar\omega_c$, we find
\begin{equation}
\frac{\hbar\omega_c}{m_e c^2}\approx 0.1 \left(\frac{r}{100 R_{\rm NS}}\right)^{-7} L^{1/2}_{ 43}\nu^{-5}_{\rm{FRB},9}{\mathcal{N}^3_{37}}.
\end{equation}
Owing to the extremely steep radial dependence, pair production rapidly becomes inefficient beyond $\sim 100 R_{\rm NS}$. As a result, at these distances, AWs seeded by FMS decay become charge-starved at amplitudes below that of the FRB signal, quenching the FMS wave decay. 
This renders the process of excitation of Alfv\'en waves irrelevant at $r\gtrsim 10^8$\,cm. However, at such large radii, the FRB amplitude exceeds $B_{\rm bg}$, if it propagates through a static dipole magnetosphere. Then, the FRB is quickly damped as described in \cite{Beloborodov2024}.

\subsection{Propagation on top of the escaping magnetic pulse}

In a more 
promising
scenario, FRBs are generated during magnetic explosions. 
Early production of an FRB then must occur inside the relativistic ejecta of the explosion -- a strong electromagnetic pulse expanding outward with speed $c$.
In this case, the ratio of the wave magnetic field to 
the pulse field remains small and
constant with radius.
Then, the
damping mechanism 
described
by \citet{Beloborodov_2022_PRL,Beloborodov2024} does not operate
while conversion into Alfv\'en waves is still possible.

The propagation of eruption-driven pulses has been studied extensively \citep[e.g.,][]{Lyubarsky_forced_reconnection2020:2001.02007v2, Beloborodov2020}. The pulse is launched at $R_0 \sim (c\mu^2/L_p)^{1/4} \sim 25R_{\rm NS}\mu^{1/2}_{33}L_{p,47}^{-1/4}$, 
where
$L_p \sim 10^{47}\,{\rm erg\,s^{-1}}$ is the pulse luminosity. The pulse sweeps up and carries plasma from the magnetosphere. The minimum number of $e^{\pm}$ pairs captured by the pulse is set by the magnetospheric density, $\mathcal{N}_{\rm min} \sim n_{\rm bg}R_0^3 \sim \mathcal{N}$. During the onset of the eruption, additional pairs are generated because of magnetic dissipation. Their density is limited by pair annihilation, $n\approx 16/(3\sigma_TR_0)$. This limit 
implies a total
number of particles $\mathcal{N}_{\rm max} \sim 4\pi n R^2_0\Delta_p  \sim 10^{41}$, where $\Delta_p\sim R_0$ is the width of the pulse. The plasma magnetization, however, remains extremely large, $\sigma_p\sim {\mathcal{E}}_p/(4\pi m_ec^2\mathcal{N})\sim 10^8{\mathcal{E}}_{ p,44}\mathcal{N}^{-1}_{41}$, with ${\mathcal{E}}_p\sim 10^{44}{\rm erg}$ representing the total electromagnetic energy in the pulse. As the pulse expands, its magnetic energy and flux are conserved; thus, its transverse dimension grows proportionally to radius, while its width stays approximately constant \citep[as shown by FFE and magnetohydrodynamic simulations of][]{Yuan_beloborodov_2022:2204.08513v2,Chatterjee2026}. Consequently, the plasma density in the pulse can be found as $n_p\sim \mathcal{N}/(4\pi \Delta_p r^2)$. As the pulse propagates outward, it quickly accelerates to high Lorentz factors, $\Gamma$. In the region of interest, the acceleration is linear, given by $\Gamma\sim r/R_0$. In the proper frame of the pulse, the plasma density, $\tilde{n}$, and magnetic field strength, $\tilde{B}$, can be determined as $\tilde{n}=n_p/\Gamma$ and $\tilde{B}=B_p/\Gamma$, where $B_p\sim \sqrt{L_p/c}/r$.


We now examine the applicability of the FFE framework for describing the decay of FRB waves in the pulse frame. First, the ratio of the plasma frequency, $\tilde{\omega}_p=\sqrt{4\pi\tilde{n}e^2/m_e}$, to the wave frequency, $\tilde{\omega}_{\rm FRB}=\omega_{\rm FRB}/\Gamma$, remains small in the inner magnetosphere, 
\begin{equation}
\tilde{\omega}_p/\tilde{\omega}_{\rm FRB}\sim 1.5\times 10^{-4}\left(\frac{R_0}{r}\right)^{1/2}\mathcal{N}^{1/2}_{41}\nu^{-1}_{\rm FRB,9}L^{3/8}_{p,47},
\end{equation} 
justifying the treatment of FRB waves as FMS perturbations. Second, the amplitude of marginally charge-starved AWs relative to the FRB stays constant with distance, 
\begin{align}
\tilde{\psi}_{\rm st}&=\tilde{E}_{\rm AW, st}/\tilde{E}_{\rm F0}\sim 8\pi \Gamma^2 \tilde{n}ecr/(\omega_{\rm FRB}(2L_{\rm FRB}/c)^{1/2})\nonumber\\& \sim 30 (L_{p,47}/L_{,43})^{1/2}\mathcal{N}_{41}\nu^{-1}_{\rm FRB,9},
\end{align}
where we used $\tilde{E}_{\rm F0}=E_{\rm F0}/\Gamma$ and $\tilde{n}=n_p/\Gamma$. Thus, the excited AWs can readily grow to amplitudes comparable to that of the FMS wave before becoming charge-starved. Therefore, three-wave decay of FRB waves propagating atop the escaping magnetic pulse proceeds within the FFE regime in the inner magnetosphere.

Finally, we estimate the number of e-foldings of the three-wave processes in the pulse, following \cite{Lyubarsky_forced_reconnection2020:2001.02007v2}. As the pulse accelerates rapidly, the maximum interaction timescale can be estimated as $t_{\rm max}\sim \min(\Delta t_{\rm FRB},r/2c\Gamma^2)=r/2c\Gamma^2$ in the region of interest. Thus,
\begin{align}
\eta
&\sim \omega_{\rm FRB}\left(\frac{B_{\rm F,0}}{B_p }\right)^2\left(\frac{r}{2\Gamma^2c}\right)\nonumber \\&\approx220\,\nu_{\rm FRB,9}L_{43}L^{-3/2}_{\rm p,47}\left(\frac{r}{30R_{\rm NS}}\right)^{-1}.
\end{align}

The FRB escapes if $\eta$ exceeds a critical value $\eta_{\rm cr}=10-30$, depending logarithmically on the level of seed Alfv\'en waves. This gives a lower limit on the emission radius of observed FRBs:
\begin{equation}
 \frac{r}{R_{\rm NS}}\gtrsim 700 \left(\frac{\eta_{\rm cr}}{10}\right)^{-1} L_{43}\nu_{\rm FRB,9}L^{-3/2}_{\rm p,47}.
\end{equation}

\subsection{Implications for kHz waves}

Finally, we consider the escape of low-frequency ($\sim\mathrm{kHz}$) waves generated by starquakes \citep[e.g.,][]{khz_blaes_1989ApJ...343..839B,khz_thompson_1996ApJ...473..322T,khz_bransgrove_2020ApJ...897..173B}. Such waves have been shown to generate ``monster'' shock waves in the magnetosphere when their amplitude approaches half the background field \citep{monster_shock_Beloborodov2022:2210.13509v3}. At this point, the $\boldsymbol{E}\times\boldsymbol{B}$ drift velocity of the plasma diverges, and plasma inertia becomes important. The front of the pulse is unaffected by nonlinear interactions and is expected to launch a shock regardless of the pulse duration. By contrast, the tail of the pulse can excite turbulence before shock formation if the number of e-foldings for the resonant processes becomes large before the shock condition, $|\boldsymbol{E}_{\rm F0}|/B_0 = 1/2$, is satisfied. This occurs when $\eta \approx \pi N_{\lambda}/2 \gg 1$, where we applied the scaling of $\gamma_{\rm NL}$ for a monochromatic wave, corresponding to $\Delta t \gtrsim 6\times10^{-4}\,\mathrm{s}\,\eta_1\,\nu_4^{-1}$, where the wave frequency is normalized to 10\,kHz. For realistic pulse durations, $\sim1\,\mathrm{ms}$, the two conditions can be satisfied in roughly the same region, leading to 
a
coupled behavior—for example, strong pair loading at the shock can modify wave interactions in the trailing part of the pulse.

In summary, we have performed three-dimensional force-free electrodynamical simulations demonstrating that mildly nonlinear FMS pulses, spanning many wavelengths, efficiently transfer their energy to trapped AWs via resonant three-wave interactions. We find that in quiescent magnetospheres of magnetars, this decay process restricts FRB escape from within $\sim 100\,R_{\rm NS}$. In a more realistic scenario---where FRBs are generated and propagate within an outward-moving magnetic pulse produced by a magnetospheric eruption---wave decay instead limits the emission region to $\gtrsim 10^3\,R_{\rm NS}L_{43}$. Overall, our results reinforce the conclusion that the escape of energetic radio pulses from the inner magnetosphere of magnetars is challenging.

\section*{Acknowledgments}
We thank Roger Blandford, Amir Levinson, Misha Medvedev, and Anatoly Spitkovsky for insightful discussions. SS is grateful to Sophia Woznichak for their help with the illustrations in the paper. This work was supported by NSF grants AST-2307395, AST-2406908 (SS and AP) and AST-2508744 (JFM), Simons Foundation (00001470, AP and SS), and facilitated by Multimessenger Plasma Physics Center (MPPC, AP and AMB), NSF grant No. PHY-2206607. A.P. additionally acknowledges support by an Alfred P. Sloan Fellowship, and a Packard Foundation Fellowship in Science and Engineering. 
AMB acknowledges support by NASA grants 21-ATP21-0056 and 80NSSC24K1229, NSF grant AST-2408199, and Simons Foundation grant 446228.
This research is part of the Frontera computing project \citep{Frontera} at the Texas Advanced Computing Center (LRAC-AST21006). Frontera is made possible by NSF award OAC-1818253. This research was supported in part by grant NSF PHY-2309135 to the Kavli Institute for Theoretical Physics (KITP).

\bibliographystyle{aasjournal}
\bibliography{apssamp}

\end{document}